\documentclass{aa}
\usepackage{graphicx}
\usepackage{txfonts}
\usepackage{color}

\begin{document}

\title{2MASSJ22560844+5954299: the newly discovered cataclysmic star with the deepest eclipse \thanks{Based on data collected with the telescopes at Rozhen National Astronomical Observatory}}
\titlerunning{2MASSJ22560844+5954299: the newly discovered cataclysmic star}
\authorrunning{Kjurkchieva et al.}

\author{D. Kjurkchieva\inst{1} \and T. Khruzina\inst{2} \and D. Dimitrov\inst{3} \and R. Groebel\inst{4} \and S. Ibryamov\inst{1,3}
\and G. Nikolov\inst{3}}

\institute{Department of Physics, Faculty of Natural Sciences, Shumen University, 115, Universitetska Str., 9712 Shumen, Bulgaria\\
  \email{d.kyurkchieva@shu-bg.net}
  \and Moscow MV Lomonosov State University, Sternberg Astronomical Institute, 13, Universitetskii pr., 119991 Moscow, Russia
  \and Institute of Astronomy and National Astronomical Observatory, Bulgarian Academy of Sciences, 72, Tsarigradsko Shose Blvd., 1784 Sofia, Bulgaria
    \and Bundesdeutsche Arbeitsgemeinschaft f\"{u}r Ver\"{a}nderliche Sterne e.V. (BAV) Munsterdamm 90, D-12169 Berlin, Germany}

\date{Received 15 March 2015 / Accepted ... 2015}


  \abstract
{The SW Sex stars are assumed to represent a distinguished
stage in CV evolution, making it especially important to study them.}
{We discovered a new cataclysmic star and carried out prolonged
and precise photometric observations, as well as medium-resolution spectral observations. Modelling these data
allowed us to determine the physical parameters and to establish
its peculiarities.}
{To obtain a light curve solution we used model whose emission
sources are a white dwarf surrounded by an accretion disk with a
hot spot, a gaseous stream near the disk's lateral side, and a
secondary star filling its Roche lobe. The obtained physical
parameters are compared with those of other SW Sex-subtype stars. }
{The newly discovered cataclysmic variable 2MASSJ22560844+5954299
shows the deepest eclipse amongst the known nova-like stars. It was
reproduced by totally covering a very luminous accretion disk by a
red secondary component. The temperature distribution of the disk
is flatter than that of steady-state disk. The target is
unusual with the combination of a low mass ratio $q \sim$~1.0 (considerably below the limit
$q = 1.2$ of stable mass transfer of CVs) and an M-star secondary.
The intensity of the observed three emission lines, H$\alpha$, He
5875, and He 6678, sharply increases around phase 0.0, accompanied
by a Doppler jump to the shorter wavelength. The absence
of eclipses of the emission lines and their single-peaked profiles
means that they originate mainly in a vertically extended hot-spot
halo. The emission H$\alpha$ line reveals S-wave wavelength
shifts with semi-amplitude of around 210 km~s$^{-1}$ and
phase lag of 0.03.}
{The non-steady-state emission of the luminous accretion
disk of 2MASSJ22560844+5954299 was attributed to the low
viscosity of the disk matter caused by its unusually high
temperature. The star shows all spectral properties of an
SW Sex variable apart from the 0.5 central absorption.}

   \maketitle
%

\section{Introduction}

The cataclysmic variables (CVs) consist of a white dwarf and a
late main-sequence (MS) star filling its Roche lobe. The white
dwarfs of the most CVs (excluding polars) are surrounded by
accretion disks. The disk and/or the hot spot are dominant sources
of their optical emission. These systems are natural laboratories
for accretion-disk physics because the timescale of their
accretion is relatively short compared to other accreting
objects. Cataclysmic variables are the closest systems whose mass
outflows are associated with accretion via a disk onto compact
objects (Noebauer et al. 2010).

The nova-like (NLs) stars are nonmagnetic cataclysmic variables
that do not show large variations like dwarf nova outbursts.
Their main photometric characteristics are small humps and reduced
flickering, while their optical spectra are dominated by wide
Balmer emission lines and HeI/HeII emission lines (for a
comprehensive description of NLs and their emission sources, see
the books of Warner (1995) and Hellier (2001)). Most nova-like
variables have orbital periods just above the period gap where
magnetic braking from the secondary star (Howell et al. 2001) is
thought to be the dominant mechanism for angular momentum loss
and to have a much stronger effect than gravitational radiation
(which is thought to be dominant below the gap). This naturally
explains the high mass-transfer rates above the gap (although one
has to keep in mind that there are also dwarf novae above the
gap). It is assumed that the NLs are in a state of ``permanent
eruption'' because of their high accretion rate, producing largely
ionized accretion disks in which the viscous-thermal instability
(driving dwarf nova limit cycles) is suppressed (Osaki 2005).
Because they stay in the high mass-transfer state for long periods
of time, they may be the closest examples of steady-state
accretion disks among the CVs.

A quarter century ago, a sub-class of NLs, called ``SW~Sex stars'',
was identified (Szkody $\&$ Piche 1990; Thorstensen et al. 1991a;
Dhillon et al. 2013) with contradictory properties: deep continuum
eclipses (with unusual V-shaped profiles) that imply a
high-inclination accretion disc and emission lines with
single-peaked profiles and shallow eclipses, instead of the
deeply eclipsing, double-peaked profiles one would expect (Horne
$\&$ Marsh 1986). Moreover, their emission lines do not reflect
the orbital motion of the white dwarf (there is a substantial
orbital phase lag of $\sim$0.2 cycle) and exhibit transient
absorption features around phase 0.5, all indicative of a complex,
possibly non-disk origin. Different mechanisms were proposed to
explain the behaviour of the SW Sex stars (Hellier $\&$ Robinson
1994; Williams 1989; Casares et al. 1996; Horne 1999; Knigge et
al. 2000). But their unusually high luminosities and accretion
geometry still do not have a satisfactory explanation (Townsley $\&$
Bildsten 2003, Townsley $\&$ Gansicke 2009, Ballouz $\&$ Sion
2009, etc.).

The recent observations revealed that the SW Sex stars represent
dominant fraction of all CVs in the orbital period range 3-4~h 
(Rodriguez-Gil et al. 2007). The prototypical NL UX UMa
occasionally also exhibits SW Sex-like behavior (Neustroev et al.
2011). Dhillon et al. (2013) propose that all NLs (excepting
VY Sct subtype) be classified as SW Sex stars. On the other
hand, models imply that CVs evolve from longer to shorter orbital
periods, driven by angular momentum loss, which means that CVs
that formed with periods more than 4~h will eventually pass
through the 3-4~h regime. Since most CVs in that range appear to
be SW Sex stars, it is reasonable to assume that systems that
evolve into that range will turn into SW Sex stars, meaning that
the SW~Sex phenomenon could be considered as an evolutionary stage
of the CV population. The exclusive evolution role makes the
SW~Sex stars important astrophysical objects. Although the
eclipses are no longer required to belong to the SW Sex subtype,
the eclipsing SW Sex systems are of particular interest, since
time-resolved observations of the eclipse provide spatial
information about the disk and physical parameters of their
configurations (Dhillon et al. 2013).

In this paper we report the results of our study of a newly
discovered SW~Sex star with deep eclipse. They are based on seven-year
photometry and medium-resolution spectroscopy covering the whole
orbital cycle.

\section{Discovery and first observations}

The photometric variability of the star GSC1 03997 00231 was
discovered in 2008 during observations of NGC 7429 by an eight-inch
telescope at a private observatory near N\"{u}rnberg (Groebel
2009). They revealed light minima with depth 0.4-0.5 mag and
duration around 1~h repeating every 5.5~h. The initial supposition
was that it is an eclipsing system consisting of two almost equal
dwarf stars with orbital period of near 11~h.

GSC1 03997 00231 seemed elongated on the first low-resolution CCD
images (as well as on the POSS and SDSS images). The GSC 2.3
catalogue gave the reason: there were two stars at this position
at 6~arcsec separation from each other: N1CB000021 (the brighter
one, 2MASS J22560768+5954286, RA = 22 56 07.68, Dec = +59 54 28.7)
and N1CB002289 (the fainter one, RA = 22 56 08.44, DEC = +59 54
29.98). The next N\"{u}rnberg observations identified the fainter
star as the variable one. Its designation is 2MASS
J22560844+5954299, so from now on we use the shorter name J2256.

J2256 was observed regularly during 2008-2009 by the eight-inch Meade
telescope and CCD camera ST6 without filter (in a semi-automated
mode) and by the ten-inch telescope with ST8 camera during 2010-2012.
The exposures were 60 s. The reduction of images (dark and
flat field corrections) were performed by the software
\emph{Muniwin} (Motl 2012). An aperture of five-pixel radius
corresponding to a 7.5 arcsec sky field was used for photometry.
The comparison stars were selected to be similar to the variable
star in magnitude and spectral type (estimated by the 2MASS (J-K)
index) and to be constant within the photometric error of 0.02
mag.

The times of the individual minima were determined by the
method of Kwee $\&$ van Woerden (1956). The linear fit of the
obtained 73 times of eclipse minima from 2008--2012 led to the
ephemeris
\begin{equation}\label{eq:1}
HJD(Min) = 2454831.4202(1) + 0.228605624(39) \times E .\\
\end{equation}

\begin{figure}
   \centering
   \includegraphics[width=0.8\columnwidth, angle=0]{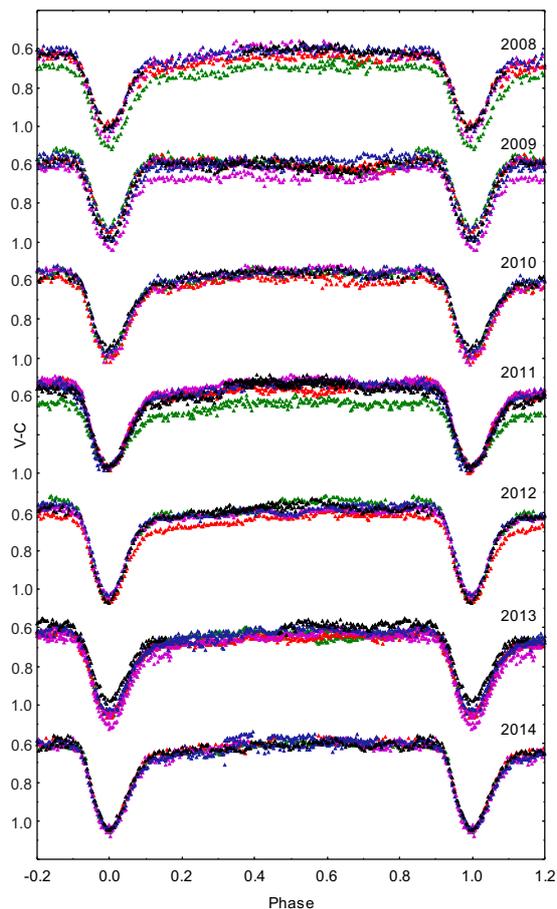}
   \caption{Representative (annual) samples of the numerous N\"{u}rnberg
   unfiltered light curves of J2256 during 2008-2014 (the symbol
   sizes correspond to the photometric error of around 0.02 mag)}
   \label{Fig1}
   \end{figure}

The intensive N\"{u}rnberg observations exhibited changes both in
the minimum depths and out-of-eclipse shape and in the asymmetry
of the eclipses (Fig. 1). These characteristics indicated a
nova-like eclipsing cataclysmic variable rather than a detached
eclipsing system. The V-shaped light minimum, as well as the
presence of ``shoulder'' on the increasing branch of some light
curves, supported this supposition (Fig. 1). The
star (IPHAS J225608.45+595430.0) in the IPHAS catalogue (of
H-alpha emission lines source) was another indication of its
classification as a cataclysmic variable (Witham et al. 2008).

To study this newly discovered cataclysmic variable we undertook
follow-up photometric and spectroscopic observations at
the Rozhen observatory.

\begin{table}
\caption{Journal of the Rozhen observations} \label{tab:1}
\centering
\begin{scriptsize}
\begin{tabular}{l c c c c c}
\hline\hline
Date       &Filter & Exp. [s]& N      & Accuracy [mag]& Telescope \\
\hline
2013 Aug 28& $I$   & 90      & 132    &  0.008 & 60 cm     \\
2013 Sept 1& $V$   & 120     & 190    &  0.015 & 60 cm    \\
2013 Sept 2& $R$   & 90      & 260    &  0.010 & 60 cm   \\
2013 Sept 4& $B$   & 120     & 200    &  0.025 & 60 cm    \\
2013 Sept 5& $I$   & 90      & 276    &  0.008 & 60 cm    \\
2013 Sept 6& $B$   & 90      & 187    &  0.01  & 2 m    \\
2013 Oct 25& spectra & 900   & 20     &        &  2 m    \\
2013 Oct 28& $B$   & 120     & 172    &  0.025 &  60 cm \\
2013 Oct 30& $V$   & 120     & 160    &  0.015 &  60 cm \\
2013 Dec 4 & $R$   & 90      & 229    &  0.010 &  60 cm \\
2014 Aug 27& $I$   & 120     & 186    &  0.008 &  60 cm \\
\hline
\end{tabular}
\label{tab1}
\end{scriptsize}
\end{table}

\begin{table*}
\caption{Coordinates and magnitudes of the comparison
stars for the data of the 60 cm telescope} \label{tab:2}
\centering
\begin{scriptsize}
\begin{tabular}{l c c c c c c}
\hline\hline
Name & RA & Dec & I (USNO-B1.0)& R (USNO-A2.0)& V (GSC 2.3.2)&  B (USNO-A2.0)\\
\hline
2MASS 22560365+5954157 & 22 56 03.66 & +59 54 15.73 & 12.73 & 14.1  & 14.51  & 16.2 \\
2MASS 22555212+5951564 & 22 55 52.12 & +59 51 56.42 & 12.26 & 13.5  & 13.92  & 15.7 \\
2MASS 22560101+5950335 & 22 56 01.02 & +59 50 33.57 & 12.34 & 13.5  & 13.19  & 14.1 \\
\hline
\end{tabular}
\end{scriptsize}
\end{table*}

\section{Rozhen observations}

\begin{table*}
\caption{Parameters of the Rozhen light curves}
\begin{center}
\begin{scriptsize}
\begin{tabular}{c r c c c c c c}
\hline \hline
Filter    & $NN$ (1)   & Date       & $t_{1}$, $t_{2}$,    & $\varphi_{1}$, $\varphi_{2}$ & $T_{min}$,  & $m_{max}$ & $m_{min}$ \\
 set      & [$NN$ (2)] &            &   JD 2456000+        &                              & JD 2456000+ &           &                     \\
\hline
$V_{1}$   & 7462.9854  & 01.09.2013 &  537.34, 537.62      & 0.310, 1.484          & 537.500630    & 14.290    & 18.259        \\
          & [-0.00074] &            &                      &                       &               &           &                 \\
$R_{1}$   & 7466.9914  & 02.09.2013 &  538.33, 538.62      & 0.660, 1.878          & 538.416432    & 14.657    & 17.478        \\
          & [4.0053]   &            &                      &                       &               &           &                    \\
$B_{1}$   & 7475.9874  & 04.09.2013 &  540.33, 540.62      & 0.391, 1.628          & 540.472960    & 14.704    & 19.036        \\
          & [13.00125] &            &                      &                       &               &           &                     \\
$I_{1}$   & 7479.9851  & 05.09.2013 &  541.32, 541.63      & 0.718, 1.967          & 541.386861    & 13.930    & 16.157        \\
          & [16.99897] &            &                      &                       &               &           &                     \\
$B_{2}$   & 7711.9829  & 28.10.2013 &  594.27, 594.52      & 0.350, 1.429          & 594.422856    & 14.791    & 18.835       \\
          & [248.99674]&            &                      &                       &               &           &                    \\
$V_{2}$   & 7720.9820  & 30.10.2013 &  596.30, 596.54      & 0.233, 1.227          & 596.480098    & 14.353    & 18.329        \\
          & [257.99583]&            &                      &                       &               &           &                    \\
$R_{2}$   & 7872.9931  & 04.12.2013 &  631.18, 631.44      & 0.798, 1.928          & 631.230704    & 14.570    & 17.520        \\
          & [410.00699]&            &                      &                       &               &           &                   \\
$I_{2}$   & 9036.9821  & 27.08.2014 &  897.29, 897.57      & 0.887, 2.051          & 897.325121    & 13.792    & 15.964        \\
          & [1573.99693]&           &                      &                       & 897.553727    &           &                    \\
\hline
\end{tabular}\label{tab3}
\end{scriptsize}
\end{center}
\end{table*}

The photometric observations of J2256 at the Rozhen National
Astronomical Observatory were carried out with (i) the 60-cm
Cassegrain telescope using the FLI PL09000 CCD camera (3056
$\times$ 3056 pixels, 12 $\mu$m/pixel, field of 27.0 $\times$ 27.0
arcmin with focal reducer), and (ii) the 2-m RCC telescope with the
CCD camera VersArray 1300B (1340 $\times$ 1300 pixels, 20
$\mu$m/pixel, diameter of field of 15 arcmin).

Table 1 presents a journal of the Rozhen observations obtained under
good atmospheric conditions (seeing 1--2 arcsec) and covering at
least a cycle. Their average accuracy (Table 1) became three to four times
bigger at the bottom of the eclipses.

The standard \emph{IDL} procedures (adapted from \emph{DAOPHOT})
were used for reducing the photometric data. All frames are
dark-frame-subtracted and flat-field-corrected. They were analyzed
using an aperture of 3 arcsec. The comparison stars (Table 2) were
selected to fulfil the requirement of being constant within 0.01 mag
during all observational runs and in all filters.

The spectral observations of J2256 were carried by the 2 m RCC
telescope equipped with the focal reducer FoReRo 2 and grism with
720 lines/mm. The resolution of the spectra is 2 pix or 2.7 $\AA$
and they cover the range 5600-7000 $\AA$. Most of the spectra
have a S/N of 16-22 excluding those at the eclipse where the S/N
value is around 7. The spectra were reduced using
\emph{IRAF} packages for bias subtraction, flat-fielding, cosmic
ray removal, and one-dimensional spectrum extraction. For
wavelength calibration, we used 30 appropriate night-sky emission
lines from the spectral atlas of Osterbrock et al. (1996). They
were fitted by third-order polynomials and the resulting dispersion
curve was with RMS$\sim$0.2 $\AA$. The spectra were normalized by
dividing by the continuum value around 6500 $\AA$.

\section{Analysis of the photometric data}

\subsection{Initial analysis}

To determine the eclipse timings of the Rozhen photometric
data and their errors, we fitted the eclipses by the function
\emph{Pearson IV} of the software {\sc Table curve 2D}
(Systat Software, Inc.). The linear fit of
all N\"{u}rnberg eclipse timings, including the later ones from
2013-2014 (Fig. 1), as well as those from the precise Rozhen
photometry, led to improving the target period and led to a new
ephemeris:
\begin{equation}
Min(HJD) = 2456537.5008(3) + 0.228605615(27) \times E
,\end{equation}
whose initial epoch corresponds to the first observed minimum at
Rozhen.

We phased the Rozhen photometric observations with the ephemeris
(2). Table 3 presents information on the sample consisting of two
light curves in each filter (indexed by ``1'' and ``2'') obtained
by the 60 cm telescope and separated by at least approximately two
months. The columns of Table 3 are as follows: set designation;
\emph{NN(1)} -- number of the orbital cycle according to the
ephemeris (1); \emph{NN(2)} -- number of the orbital cycle
according to the ephemeris (2); date; $t_1$, $t_2$ -- times of the
beginning and end of the observations; $\varphi_1$, $\varphi_2$ --
phases of the beginning and end of the observations according to
(2); T$_{min}$ -- time of the observed minimum; $m_{max}$ and
$m_{min}$ -- visual magnitudes corresponding to the maximum and
minimum brightness during the corresponding set of observations.

\begin{table}
\caption{Depths and widths of the eclipse} \label{tab:4}
\centering
\begin{scriptsize}
\begin{tabular}{c c c}
\hline\hline
Filter& Depth    & Width \\
      & mag      & phase units \\
\hline
$B$ & 4.0--4.3   & 0.220  \\
$V$ & $\sim$ 4.0 & 0.210 \\
$R$ & 2.8--3.0   & 0.225 \\
$I$ & $\sim$ 2.2 & 0.235 \\
\hline
\end{tabular}
\end{scriptsize}
\end{table}

The qualitative analysis of the Rozhen\emph{ BVRI} light curves
(Fig. 2) led us to several conclusions. (i) There are changes of
the light curves of the two filter sets by $\sim$0.06--0.14 mag
(bigger in \emph{V} and \emph{I} bands). (ii) The eclipse depths
of the Rozhen data are considerably bigger (by a factor of 4--8)
than those of the N\"{u}rnberg ones (Fig. 1). This effect is due
to the low spatial resolution of the N\"{u}rnberg
observations, leading to considerable light contribution of the
neighbouring star and thus to reducing of the true amplitude of
variability. (iii) There is no pre-eclipse hump, except a light
curve $B_1$ with a weak hump (Fig. 2). (iv) The light level after
the eclipse is lower than the level before it. (v) There is a prolonged
plateau outside the eclipse in filters \emph{B} and \emph{V}.

 \begin{figure*}
   \centering
   \includegraphics[width=16cm, angle=0]{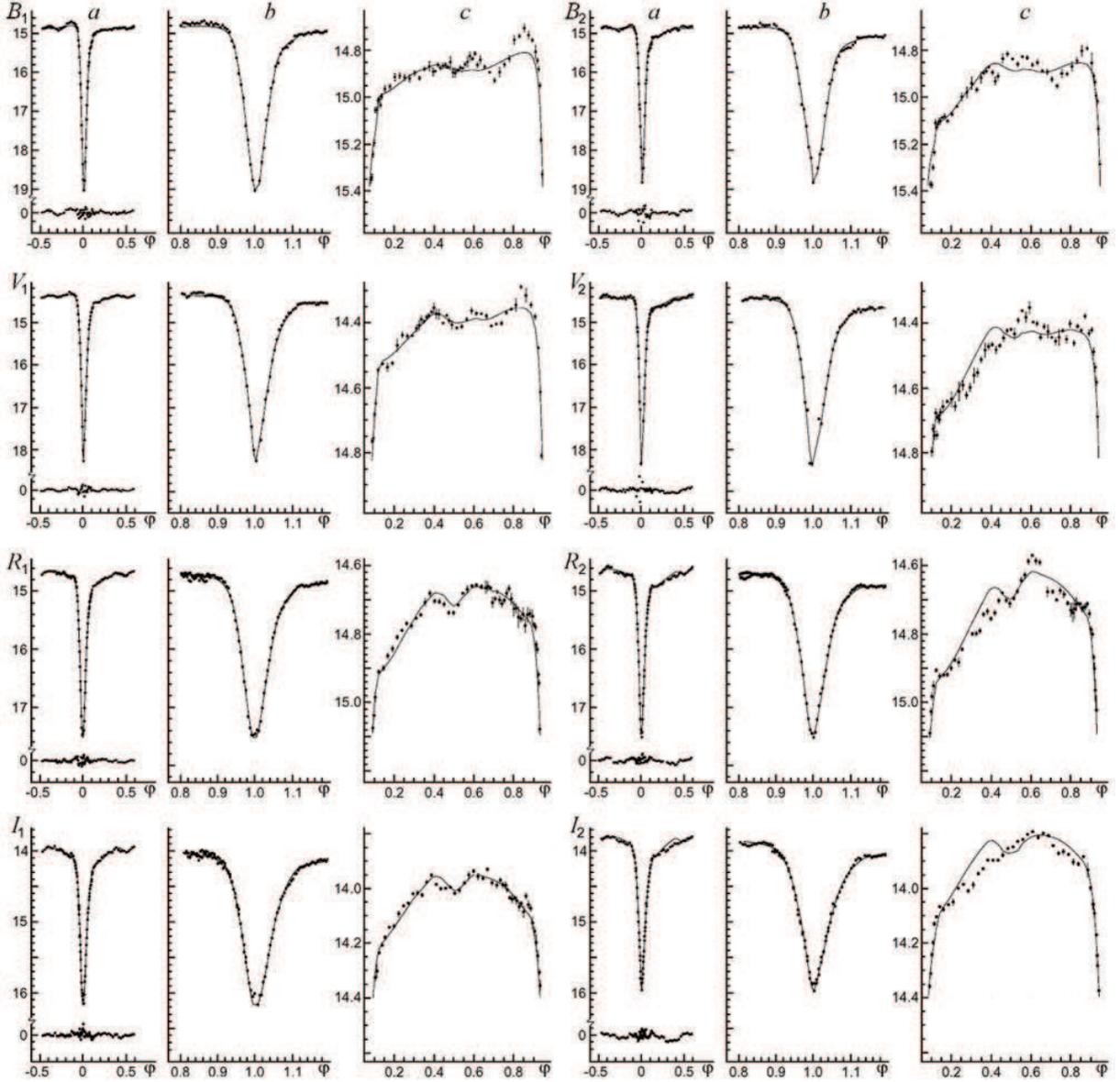}
   \caption{(a) Rozhen photometric data of J2256 (points with the corresponding error bars),
the synthetic light curves (continuous lines), and the corresponding residuals; (b) details around the eclipse
minima (the error bars are smaller than the symbol size); (c)
details of the out-of-eclipse parts of light curves. }
   \label{Fig2}
   \end{figure*}

The most distinguishable feature of J2256 is its deep eclipse:
$\sim$4 mag in $V$. The eclipse depths of the majority of the
eclipsing nova-like variables are $\leq$1 mag. The deepest
eclipses belong to the SW Sex stars. Only nine SW Sex stars in the
list of Rodriguez-Gil et al. (2007) have eclipse depths more than
2.0 mag (Khruzina et al. 2013). Two SW Sex stars, DW UMa
(Stanishev et al. 2004) and V1315 Aql (Papadaki et al. 2009),
normally have eclipse depths less than 2.0 mag, but on rare occasions
and only within two to three days do they show deeper eclipses of 3.2-3.4 mag.
Dimitrov $\&$ Kjurkchieva (2012) report the SW Sex star 2MASS
J01074282+4845188 with an eclipse depth of V$\sim$2.9 mag, but the
newly discovered cataclysmic star J2256 turned out to have an even
deeper eclipse. In fact, it shows the deepest eclipse among the
known eclipsing nova-like variables.

\subsection{Light curve solution}

The shapes of the J2256 light curves are typical of nova-like CVs.
Particularly, 2MASS J01074282 + 4845188, observed by us with the
same equipment, has a similar light curve (Khruzina et al. 2013).
That is why we used the model applied successfully for other
cataclysmic variables to determine the system parameters of J2256.
A detailed description is given in Khruzina (1998, 2011). The
configuration consists of a white dwarf and late-type secondary
filling of its Roche lobe. The shape of the secondary is
determined by the Roche potential. The effects of gravitational
darkening and non-linear limb darkening are taken into account
when calculating the emission of its surface elements. The
emission of the white dwarf with radius $R_{wd}$ is assumed for
the black-body type that corresponds to effective temperature
$T_{wd}$. It is surrounded by an opaque, slightly elliptical
accretion disk, coplanar to the orbital plane of the system. The
outer (lateral) surface of the unperturbed disk presents part of
ellipsoid with semi-axes $a, b, c$, while its inner surfaces are
parts of two paraboloids with parameter $A$. The thickness of the
disk outer edge $\beta_d$ is specified by the parameters $a$ and
$A$. Another geometrical parameter $\xi (q)$, which was used in
the solution, represents the distance between the centre of the
white dwarf and L1. The collision of the gas stream from the
secondary with the rotating disk results in formation of hot spot
described by a half ellipse. All linear sizes of the model are in
units of the binary separation $a_{0}$. Furthermore, we use the
unit $\xi$ to express the disk radius in order to compare it to
similar structures in other CVs (see Section 4.3). Moreover, $\xi$
turns out to be a useful parameter at the first stage of the light
curve solution when the mass ratio is unknown, and $\xi$ serves to
set the possible ranges for varying $R_{wd}$ and $R_{d}$. (If they
were in $a_{0}$ there is a considerable probability the component
sizes to exceed the Roche lobe that will cause stop of the
calculations.)

The fluxes from the differential emitting areas are in conditional
units because the Planck function gives the flux from
1 cm$^2$ per unit wavelength range (in our case, centimetre), while
the linear measure of the code for the light curve synthesis is
the distance $a_0$, whose value is not known in advance.

The temperature profile of the disk is determined by the parameter
$\alpha_{\rm{g}}$
\begin{equation}
T_{d}(r) = T_{in} \left(\frac {R_{in}}{r}\right)^{\alpha_g}
\end{equation}
($\alpha_{\rm{g}}$=0.75 corresponds to the equilibrium state of
the disk).

While the traditional models of CVs assume that the gas
stream is transparent to the emission of the shock wave, and the
heated region, called a ``hot spot'', is the area of the
interaction of the stream with the disk, the gas stream is opaque
in our model. As a result, the heated region, analogue of the
``classical'' hot spot, consists of two components (Fig. 3): part
of the hot line near the disk and disk region at the leeward side
of the hot line.

The optically opaque part of the gas stream, the hot line, is
approximated by part of the elongated ellipsoid with axes $a_v, b_v,
c_v$.  Its major axis $b_v$ coincides with the axis of the gas
stream from the inner Lagrange point L1. The temperature of the
hot-line surface changes according to a cosine law (for details,
see Khruzina, 2011):
\begin{equation}
T_{hl}(r) = T_d(r) +\Delta T_{n,max}\cos (\delta r)
,\end{equation}
where $\Delta T_{ww,max}$ and $\Delta T_{lw,max}$ are the maximum
corrections to the disk temperature $T_d(r)$ corresponding to the
windward and leeward sides near the disk, $\delta r$ is the
distance of the current hot-line point from the disk edge, and $n$ is
for $ww$ or $lw$. The local temperature at the disk area inside the
hot spot varies by a cosine law with the maximum value at the point U
of the intersection between the hot-line axis and the disk.

The adjustable parameters of the light curve solution were: (a)
parameters of the stellar components: mass ratio $q =
M_{wd}/M_{red}$; orbital inclination $i$; effective temperatures
of the white dwarf $T_{wd}$ and red star $T_{red}$; radius of the
white dwarf $R_{wd}$; (b) parameters of the accretion disk:
temperature $T_{in}$ of the inner region (boundary layer);
eccentricity $e$ ($e \leq$ 0.1); large semiaxis $a$; parameter
$\alpha _g$ defining the temperature distribution along its
radius; azimuth of the disk periastron $\alpha _e$; parameter of
the paraboloid surface \emph{A} (the thickness of the outer disk
edge $\beta_d$ = f({\it A, a}) is calculated); (c) parameters of
the hot line: semiaxes of the ellipsoid $a_v, b_v, c_v$; maximum
temperature corrections $\Delta T_{ww,max}$ and $\Delta
T_{lw,max}$; (d) parameters of the hot spot: total size
$R_{sp} = R_{hl}+R_{hs}$ in units $a_{0}$ ($R_{hl}$ is the radius
of the hot line in the orbital plane and $R_{hs}$  the size of the
hot-spot part that is not covered by the hot line in the orbital
plane, Fig. 3); two additional parameters if the hot spot radius is
bigger than the disk thickness (Khruzina 2011). In fact, the
hot-spot centre coincides with the hot-line centre, and the hot
spot presents a semi-ellipse at the outer disk surface from the
leeward side of the hot line. The linear size $R_{sp}$ is the
distance in the orbital plane between the hot-line centre
and the hot-spot periphery. The angular hot-spot size is
correspondingly $\Delta \beta = 360 \times R_{sp} / (2 \pi R_d )$.
The radius $R_{hl}$ is quite small for the most CVs
because the gas stream is narrow near the disk, but J2256 is not
such a case (see further).

Besides the mentioned 17(19) parameters, there are several
technical model parameters: phase corrections $\Delta \varphi$
(chosen to make the middle of the white dwarf eclipse to be at
phase zero) and parameter $F_{opt}$ for normalizing the fluxes
of each filter.
Any additional information about the system could be used to
reduce the number of free parameters or at least to limit the
range of possible values of varying so many parameters in the
procedure of the light curve solution. In our case we used the
consideration that the modelled light curves of J2256 are obtained
by the same equipment and by the same reducing procedure (Table
3). This requires the same energy unit to be used for transfer of
the synthetic fluxes into visual magnitudes of the different light
curves in the same filter (Khruzina et al. 2003). In the opposite
case, different energy units will be necessary for different
equipment suitable for their inherent characteristics (for
instance, the different widths and central wavelengths of the
filters cause differences in the corresponding light curves).

We used the Nelder-Mead method for the light curve
solution (Press et al. 1986). Owing to the large number of
independent variables, there was a set of local minima in the
multi-dimensional space of parameters, and we used dozens of
different initial approximations to search for the global minimum
for each light curve. The estimate of the fit quality was the
expression
\begin{eqnarray}
\chi^2=\sum^N_{j=1}\frac{(m_j^{theor}-m_j^{obs})^2}{\sigma^2_j},
\end{eqnarray}
where $m_j^{theor}$ and $m_j^{obs}$ are the theoretical and
observed magnitudes of the star at orbital phase $\varphi_j$,
$\sigma^2_j$ is the dispersion of the observations at orbital
phase $\varphi_j$, and $N$  the number of normal points in the
curve.

The light curve solution was carried out by the following procedure.
To remove the random light fluctuations, we performed preliminary
averaging of the Rozhen data within the phase range of 0.01
(excluding the eclipse) for each band.

The fluxes $ F(\varphi)$ of all system components were
calculated by Planck function in conditional units $ F(\varphi) =
F_{wd}(\varphi) + F_{red}(\varphi) + F_d(\varphi) +
F_{hl}(\varphi),$ and their normalization was made at phase
$\varphi$ = 0.25 of each curve by the formula $m(\varphi) -
m(0.25) = -2.5 \log [F(\varphi) / F(0.25)]$. At the beginning of
the first stage, we varied the basic system parameters freely
(whose values are the same for both light curves in each filter)
for each fixed value of $q$ from the range 0.4--9.2 with step 0.1.
The ranges of the basic parameters were \emph{i} = 40--90$^0$,
$T_{wd}$ = 8000--50000 K, $T_{red}$ = 2000--10000 K, $R_{wd} /\xi$
= 0.01--0.09, and $R_d ({\rm max}) / \xi$ = 0.35--0.90. By this
procedure we reduced the acceptable range of \emph{q} to 0.9--1.1.
After that we again varied the basic parameters plus $q$
(in the newly-obtained range) freely and obtained best solution for each
curve. As a result we calculated the averaged values of the basic
parameters (from their values for the all curves): \emph{q} =
1.004; \emph{i} = 78.8$^\circ$, $R_{wd} /\xi$ = 0.0227; $T_{wd}$ =
22720 K. Furthermore, we fixed the basic parameters and varied the rest
model parameters plus normalization parameter $F_{opt}$. Its
values correspond to the same magnitudes $m_{opt}$ for each filter
by $m(\varphi) - m_{opt} = -2.5 \log [F(\varphi))/ F_{opt}]$.
Then, $F_{opt}$ was calculated as an averaged value from the two
sets for each filter.

The goal of the second stage of the procedure was to search for
the best solution for all eight light curves taking
the normalization of the fluxes to $F_{opt}$  into account for a given filter
and fixed values of the basic parameters. The temperature
$T_{red}$ was not kept fixed because of the significant differences in
its values as obtained during the first stage. The final light curve
solution was obtained by varying the remaining parameters to
search for the best fit ($\chi^{2}_{min}$). The obtained
parameters are given in Table 5. Figure 2 shows our
synthetic light curves, the observational data, and the
corresponding residuals, while Figure 3 exhibits the geometry of
the J2256 configuration.

\begin{figure}
   \centering
   \includegraphics[width=9cm, angle=0]{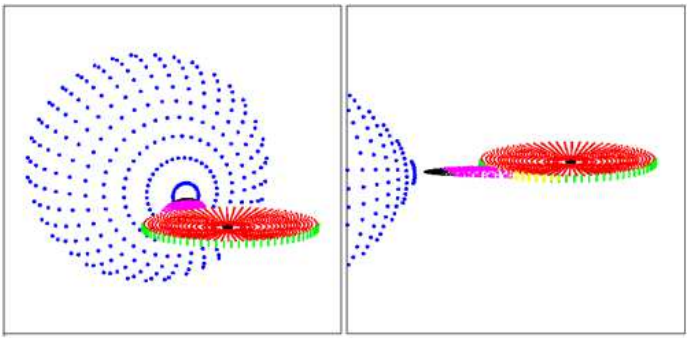}
\caption{3D configuration of J2256 at phases 0.54 (left) and 0.80
(right). Its components are shown by different colours:
blue -- secondary star; black point -- white dwarf; red and green
-- inner and lateral parts of the disk; yellow - hot-spot part that is not covered by the hot line; magenta -- heated part of
the gas stream and black -- its cold part. }
   \label{Fig3}
   \end{figure}

To estimate the errors of the adjusted parameters, we used the
procedure described in Khruzina et al. (2013). The very
small errors of the Rozhen photometric data of J2256 meant that the residuals
of all obtained solutions were bigger than the critical value
$\chi^{2}_{\alpha, N}$. Usually, the value of the probability of
rejecting the correct solution is chosen to be the level of significance $\alpha$=0.001.
Thus, for our light curves with N = 68--89, we obtained
$\chi^{2}_{0.001, N}$ = 111-138. That is why we
estimated the impact of the changing of a given parameter on the
solution quality. For this aim we varied each parameter around its
final value until reaching (for instance) level
1.1$\chi^{2}_{min}$ ($\chi^{2}_{min}$ is given in Table 5), whereas
all remaining parameters are kept equal to their final values. The
difference between the newly obtained parameter value and its
final value (Table 5) determines the corresponding error. The
parameter errors obtained in this first method are given
in Table 5 (the numbers in brackets).

A second, very time-consuming, approach to estimating the error
of each parameter or the stability of the solution is based
on arbitrary (not fixed) values of all the other parameters.
For instance, to determine the precision of \emph{q} one should
vary the all rest parameters for each value of \emph{q }(by
Nelder-Mead method, see Fig. 4). The error values estimated by the
second approach are an order bigger than those of the first method
(for fixed rest parameters), namely: $\Delta q \approx$ 0.1,
$\Delta i \approx 1.1^0$, $\Delta R_{red}/a_0 \approx$ 0.009,
$\Delta \xi /a_0 \approx$ 0.011, $\Delta R_{wd}/a_0 \approx$
0.0035, $\Delta T_{wd} \approx$ 1500 K, and $\Delta T_{red} \approx$
80 K. Usually, this second approach to estimating the
parameter errors is applied for tasks with a small number of
parameters or only for basic parameters.

\begin{figure}
   \centering
   \includegraphics[width=6.5cm, angle=0]{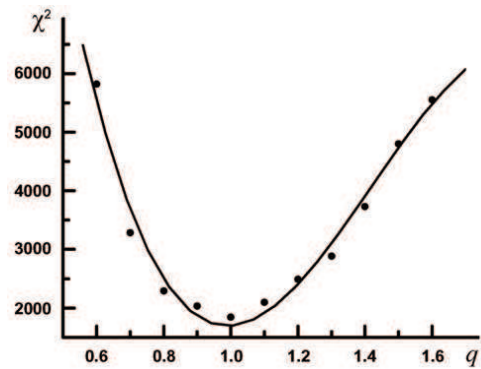}
   \caption{Illustration of the \emph{q}-search method for the light curve $V_1$}
   \label{Fig4}
   \end{figure}

Reproducing the observed light curves by the synthetic ones is
good for the eclipses (Fig. 2), while it is acceptable for the
out-of-eclipse parts. The code does not manage to describe all
small features precisely, including pre-eclipse hump in $B_1, B_2,
V_1$. Some of these discrepancies may be attributed to the
flickering that is stronger in the bluer filters, but most of them
are probably result of deficiencies in our model: (i) it does not
take geometric inhomogeneities of the accretion disk and possible
implicit dependence of some model parameters into account; (ii) it
neglects light contribution of the interstellar envelope, assuming
that it has small density; (iii) the model does not take the light
from the stream near L1 into account, assuming that its low
temperature means low radiation in the optical range. Thus, the
obtained parameter set (Table 5) should be considered as a
reasonably good (but not perfect) solution to the complex
multi-parametric inverse problem to reproduce numerous
multi-colour data stemming from a complex multi-component
configuration.

\begin{table*}
\caption{Parameters of the light curve solution of J2256 in
2013--2014} \label{tab:5}
\centering
\begin{scriptsize}
\begin{tabular}{lcccccccc}
\hline \hline Parameter  & $B_1$ & $B_2$ & $V_1$ & $V_2$ & $R_1$ &
$R_2$ & $I_1$ & $I_2$\\
\hline JD~2456000+& 540   & 594   & 537 & 596   & 538   & 631   &
541   & 897  \\ \hline $N$        & 77 & 72    & 68    & 75    &
87    & 88    & 89    & 80   \\ \hline
$q=M_{wd}/M_{red}$  &\multicolumn{8}{c}{1.004 $\pm$ 0.013}\\
$i$, $^\circ$& \multicolumn{8}{c}{78.8 $\pm$ 0.1 }\\
$<R_{red}>$, $a_0$& \multicolumn{8}{c}{0.3903(14)} \\
$\xi$, $a_0$& \multicolumn{8}{c}{0.5004(13)} \\
$T_{red}$, K   &  3151 $\pm$ 78& 3488 $\pm$ 56 & 3187 $\pm$ 27 & 3143 $\pm$ 51 & 3183 $\pm$ 18 & 3208 $\pm$ 24 & 3120 $\pm$ 9 & 3200 $\pm$ 24\\
$m_{opt}$ & \multicolumn{2}{c}{14.9}&\multicolumn{2}{c}{14.4} &\multicolumn{2}{c}{14.9} &\multicolumn{2}{c}{14.4} \\
$F_{opt}$, conditional units &\multicolumn{2}{c}{34.07}&\multicolumn{2}{c}{26.38} &\multicolumn{2}{c}{14.96}&\multicolumn{2}{c}{7.929}\\
\hline \multicolumn{9}{c}{White dwarf}\\ \hline
$R_{wd}$   &\multicolumn{8}{c}{0.0227(2)$\xi$ = 0.0114(1)$a_0$} \\
$T_{wd}$, K  &\multicolumn{8}{c}{22720 $\pm$ 110} \\
\hline \multicolumn{9}{c}{Accretion disk} \\ \hline
$e$            &    0.004(1) &  0.006(4) &  0.074(6) &  0.058(5) & 0.044(6) &  0.055(3) &  0.076(1) &  0.090(6) \\
$R_d$, $\xi$   &    0.646(1) &  0.649(7) &  0.663(3) &  0.637(10)& 0.576(6) &  0.608(2) &  0.594(2) &  0.715(11) \\
$R_d/a_0$      &    0.3218   &  0.323    &  0.309    &  0.302    & 0.276    &  0.288    &  0.276    &  0.328     \\
$a$, $a_0$     &    0.3218(3)&  0.323(3) &  0.309(1) &  0.302(4) & 0.276(3) &  0.288(1) &  0.276(9) &  0.328(5) \\
$0.5\beta_d$,$^\circ$ & 0.8(1) & 0.8(1)&    1.3(1)   &  1.4(1) & 2.1(1) &  2.1(1) &  1.8(1)   &  1.3(1) \\
$\alpha_e$, $^\circ$  & 129.5(2) &  116(2) & 173(1) &   166(2) &121(2) &  112(2) &  174(1)   &  160(4) \\
$\alpha_g$     &    0.543(3) &  0.583(2) &  0.532(2) &  0.564(3) & 0.492(2) &  0.511(2) &  0.496(2) &  0.485(3) \\
$<T_{in}>$, K  &38180 $\pm$ 180& 41834 $\pm$ 228& 36215 $\pm$ 165& 39293 $\pm$ 220 & 36395 $\pm$ 154&38064 $\pm$ 217&35839 $\pm$ 166&37578 $\pm$ 215\\
\hline \multicolumn{9}{c}{Hot spot} \\
\hline
$R_{hl}/a_0$ & 0.069 &  0.0613 &  0.1191 &  0.104 &  0.1136 &  0.153 &  0.139 &  0.161\\
$R_{hl}$, $^\circ$   & 12.3  &   10.9  & 22.1    &  19.7  & 23.6    &  30.4  & 28.9   &  28.1 \\
$R_{hs}/a_0$ & 0.159 &  0.1755 &  0.226  &  0.192 &  0.122  &  0.150 &  0.232 &  0.3095 \\
$R_{hs}$, $^\circ$   & 28.3  &  31.1   & 41.9    &  36.4  & 25.3    &  29.8  & 48.2   &  54.1 \\
$R_{sp}/a_0 = (R_{hl}+R_{hs})/a_0$&0.228(13) & 0.237(1) & 0.346(11) & 0.297(14) & 0.236(22) & 0.303(26) & 0.371(9) & 0.47(2)\\
$R_{sp}/a_0 = (R_{hl}+R_{hs})$, $^\circ$ & 40.6 & 42.0 & 64.2 & 56.3 & 49.0 & 60.3 & 77.0 & 82.1\\
$T_U$, K &  11325 $\pm$ 860& 12590 $\pm$ 1115& 17378 $\pm$ 1235& 17322 $\pm$ 1170 & 17949 $\pm$ 1600&22138 $\pm$ 1554&18992 $\pm$ 374&24880 $\pm$ 1570\\
$0.5z_{sp}$, $^\circ$ & 1.3(2) &    1.1(1) & 1.4(1) &   1.6(2) & 2.3(2) &    2.0(1) &    1.9(1) &    1.5(1) \\
\hline \multicolumn{9}{c}{Hot line} \\
\hline
$a_v$, $a_0$ &  0.0347(1) & 0.0221(4) & 0.070(3)  & 0.0521(2)& 0.084(4)  & 0.117(6)  & 0.089(1)  & 0.103(2)\\
$b_v$, $a_0$ &  0.379(1)  & 0.348(10) & 0.285(10) & 0.308(6) & 0.415(10) & 0.416(10) & 0.408(7)  & 0.367(5) \\
$c_v$, $a_0$ &  0.0045(5) & 0.0047(3) & 0.0083(5) & 0.009(1) & 0.0114(7) & 0.0109(3) & 0.0088(2) & 0.0092(5) \\
$T_{ww,{\rm max}}$, K &17330 $\pm$ 1960&16755 $\pm$ 1350& 15305 $\pm$ 1993& 14580 $\pm$ 1882 & 16762 $\pm$ 1100&14092 $\pm$ 1088&15199 $\pm$ 1173&16026 $\pm$ 980\\
$T_{lw,{\rm max}}$, K &16890 $\pm$ 370 & 17468 $\pm$ 1009& 20291 $\pm$ 992 & 20817 $\pm$ 997 & 19959 $\pm$ 1068&21641 $\pm$ 1024&22696 $\pm$ 309 &27397 $\pm$ 1293\\
$\beta_1$, $^\circ$&    31.5(1) &   35.2(4) &   26.5(2) &   28.3(1) & 16.4(2) &   13.6(5) &   19.7(1) &   23.5(5)\\
\hline $\chi^{2}_{min}$ &  1525 & 2267 & 2313 & 2710 & 1318 & 2252 & 1336 & 3636 \\
\hline
\end{tabular}
\end{scriptsize}
\begin{footnotesize}
\begin{flushleft}
Legend: $N$ -- number of average points of the light curve; $R_d$ -- radius of the disk at apoastron in units $\xi$ (distance between the center of the white dwarf and L1) and in units $a_0$; $a$, $a_0$ -- main semi-axis of the disk in units of binary separation $a_0$; 0.5$\beta_d$ -- half thickness of the outer disk edge; <T$_{in}$> -- average temperatures of the internal disk region (boundary layer); $R_{hl}$ -- radius of the hot line (in the orbital plane) in units $a_0$ and degrees; $R_{hs}$ -- size of the hot-spot part that is not covered by the hot line (in the orbital plane) in units $a_0$ and degrees; $R_{sp}$ -- total radius of the hot spot in units $a_0$ and degrees; $T_U$ -- temperature of the gas stream axis; 0.5$z_{sp}$ -- half thickness of the hot spot; $a_v, b_v, c_v$ -- hot line axes; $T_{ww, max}$ and $T_{lw, max}$ -- temperatures $T_{hl}$ of the windward and leeward sides of the gas stream near the disk ($r = 0$) according to (4); $\beta_1$ -- the angle between the axis of the gas stream
and the line connecting the mass centers of the components.
\end{flushleft}
\end{footnotesize}
\end{table*}

\begin{figure}
   \centering
   \includegraphics[width=9cm, angle=0]{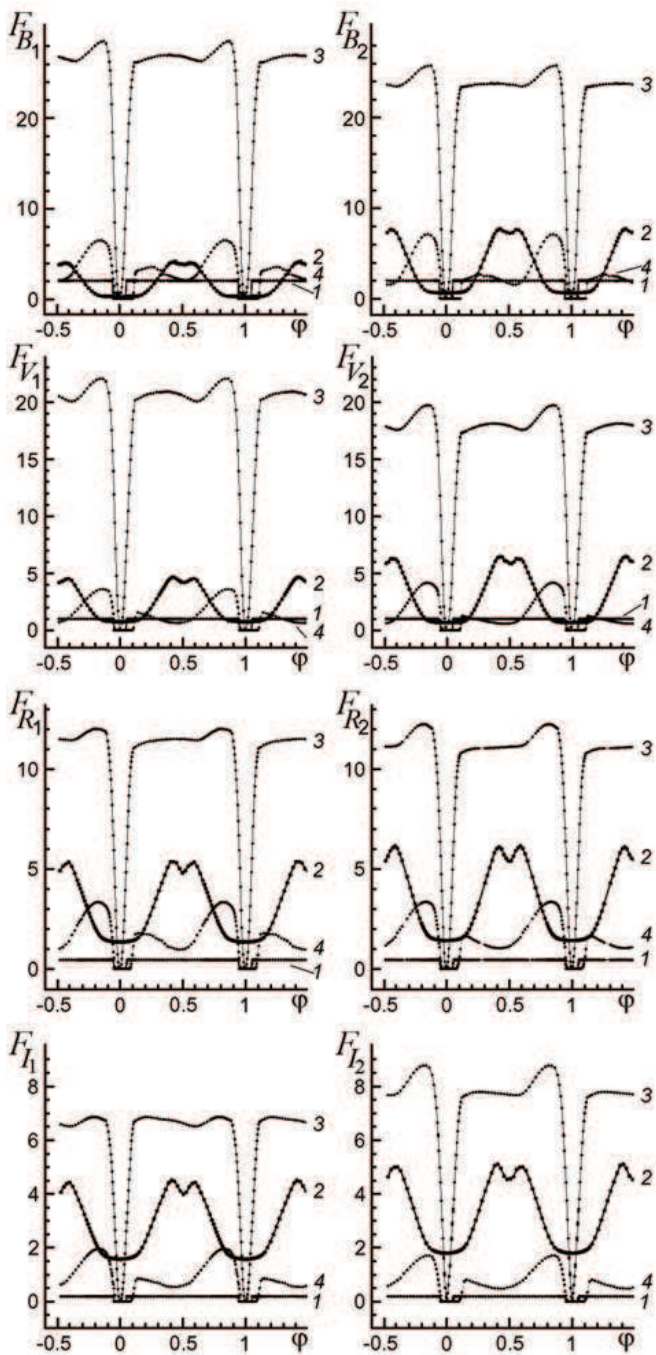}
   \caption{Light contribution of the different energy sources of the
system J2256 (in conditional units): 1 -- white dwarf, 2 -- normal
star, 3 -- accretion disk with a surface hot spot, 4 -- hot line}
   \label{Fig5}
   \end{figure}

\begin{table*}
\caption{Relative fluxes (in percentages) from the
different emission components of J2256} \label{tab:6}
\begin{center}
\begin{scriptsize}
\bigskip
\bigskip
\noindent\begin{tabular}{lcccccccc} \hline\hline
Parameter/Set & $B_1$ & $B_2$ & $V_1$ & $V_2$ & $R_1$ & $R_2$ & $I_1$ & $I_2$\\
\hline $^*F_{\rm max,total}$  &37.1  & 35.6 & 27.5 & 26.0 & 18.7 &  19.4 & 12.0 & 13.7 \\
\hline
\multicolumn{9}{c}{White dwarf} \\
$F_{\rm min}$($\varphi$=0.0)&   \multicolumn{2}{c}{0.0}&\multicolumn{2}{c}{0.0}&\multicolumn{2}{c}{0.0}&\multicolumn{2}{c}{0.0}\\
$^{**}\varphi_{ecl}$ & \multicolumn{2}{c}{0.0435}&\multicolumn{2}{c}{0.0466}&\multicolumn{2}{c}{0.0495}&\multicolumn{2}{c}{0.0460}\\
$F_{\rm max}$ & 5.3 & 5.5 & 3.6 & 3.7 & 2.4 & 2.3 & 1.5 & 1.3 \\
\multicolumn{9}{c}{Secondary star} \\
$F_{\rm min}$($\varphi$=0.0)&   0.6  & 1.8  & 2.7  & 2.5  & 7.1  & 7.3  &13.1  &13.1  \\
$F_{\rm max}$($\varphi$=0.5)&  10.1  &20.3  &15.2  &22.5  &25.5  &27.7  &33.2  &33.1  \\
Quadratures                 &  11.1  &21.5  &16.8  &24.6  &28.6  &31.5  &37.4  &36.9  \\
\multicolumn{9}{c}{Disk with hot spot} \\
$F_{\rm min}$($\varphi$=0.0)&          0.2  & 0.2  & 0.1  & 0.1  & 0.2  & 0.0  & 0.0 & 0.7  \\
<$F_{plato}$>($\varphi\sim$0.1--0.6)&  72.0 & 66.6 & 75.6 & 68.8 & 61.5 & 56.7 &56.7 &56.9 \\
Orbital hump ($\varphi\sim$0.8)&       76.5 & 72.2 & 80.0 & 75.8 & 64.2 & 62.9 &57.5 &64.2 \\
\multicolumn{9}{c}{Hot line} \\
$F_{\rm min}$($\varphi$=0.0)&   0.0 &   0.0 &   0.0 &   0.0 &   0.0 &   0.0 &   0.0 &   0.0 \\
$F_{\rm max}$($\varphi$=0.2)&   9.4 &   7.3 &   5.8 &   4.6 &   9.6 &   8.2 &   6.7 &   5.8 \\
$F_{\rm max}$($\varphi$=0.8)&  17.3 &  19.9 &  13.1 &  15.8 &  17.6 &  17.0 &  16.7 &  12.4 \\
$^{**}\varphi_{ecl}$ & 0.103 & 0.109 & 0.099   & 0.106 & 0.082 &   0.078   & 0.085 & 0.098 \\
\hline
\end{tabular}
\end{scriptsize}
\end{center}
\begin{footnotesize}
Note: $^*F_{\rm max,total}$ is in conditional units; $^{**}\varphi_{ecl}$ is the orbital phase of the appearing of
the corresponding component after its eclipse.
\end{footnotesize}
\bigskip
\end{table*}

\subsection{Analysis of the results}

The analysis of the system parameters of J2256 (Table 5)
allowed us to established which characteristics are typical of
eclipsing CVs of SW Sex subtype and which are peculiar.

(1) The emission of J2256 is dominated by the accretion disk with
the hot spot whose relative contributions are
$\sim$67--77 $\%$, $\sim$69--80 $\%$, $\sim$57--64 $\%$, and
$\sim$57--64 $\%$ in the \emph{B, V, R, I} bands, respectively (Table 6, Fig. 5).
The biggest relative contribution of the hot spot around phases
0.7--0.85 is $\sim$3--6 $\%$, i.e. considerably smaller than that
of the disk itself. The relative contributions of the secondary
star are $\sim$10-21 $\%$, $\sim$15--25 $\%$, $\sim$26--32 $\%$,
and $\sim$33--37 $\%$ in the \emph{B, V, R, I} bands, respectively. The small (11
$\%$) contribution of the secondary in the set $B_1$ explains the
visibility of the small pre-eclipse hump on the corresponding
light curve. The bigger (21 $\%$) contribution of the secondary to
the curve $B_2$ is attributed to its higher temperature. The light
curve of the secondary contribution (Fig. 5) does not correspond
to an ellipsoidal variability but rather to a ``reflection
effect'' that indicates a significant heating of its surface
facing to the white dwarf by the hot radiation from the disk
boundary layer ($T_{in}\sim$ 36000--43000 K). The third light
contribution belongs to the hot line: biggest in the \emph{B} band
($\sim$7--20 $\%$) and smallest in the \emph{I} band ($\sim$6--17
$\%$). As one expects, the white dwarf makes small contribution to
the emission of J2256: $\sim$5.4 $\%$, $\sim$3.6 $\%$, $\sim$2.4
$\%$, and $\sim$1.4 $\%$ in filters \emph{B, V, R}, and \emph{I}, respectively.

(2) The size of the accretion disk $a/a_0 \sim 0.3$ (Table 5) is
relatively small (Fig. 3) compared with that of the secondary
component, and the disk eclipse is total. But the disk of J2256 has
a relative radius  $R_d \sim (0.6-0.7) \xi$ that is similar to those
of two other SW Sex stars: $R_d >0.6 L1$ for SW Sex (Dhillon et
al. 1997) and $R_d = 0.8 L1$ for DW UMa (Dhillon et al. 2013). The
obtained value $\beta_{\rm{d}} \sim 2-4 \degr$ means that the disk
of J2256 is relatively thin. The radius of the J2256 hot
line near the disk is 11--28$^o$,  while the angular size of the
uncovered part of the hot spot is 25--54$^o$ (Table 5).

(3) The temperature $T_{\rm{in}}$ of the boundary layer of the
disk exceeds that of the white dwarf $T_{\rm{wd}}$ by factors of
1.5--1.7. The values of $T_{\rm{in}}$ in all filters are well
above the standard temperatures for cataclysmic variables at
quiescence (10000--30000 K) but are similar to those during their
outbursts (Horne $\&$ Cook 1985).

(4) The value $\alpha_{\rm{g}}\approx$ 0.5 of the parameter
determining the temperature distribution along the disk radius is
typical of cataclysmic variables after outbursts or in
an intermediate state. The $\alpha_{\rm{g}}$ value of J2256 is
less than that of UX~UMa (0.6) but bigger than for 2MASS
J01074282+4845188 (0.2). The deviation of the emission of the
luminous accretion disk of J2256 from the steady-state disk
emissions of NLs (with $\alpha_{\rm{g}}\approx$ 0.75) may be
attributed to the low viscosity of the disk matter caused by its
unusually high temperature.

For comparison, Rutten et al. (1992) studied accretion disks of
six nova-like CVs (RW Tri, UX UMa, SW Sex, LX Ser, V1315 Aql, V363
Aur) and found that the temperatures of their disks close to the
white dwarf are 10000--30000 K, and the temperatures of the hot
spots are higher than the disk by a few thousand K. SW Sex and
V1315 Aql, which are SW Sex-subtypes, exhibit a flatter
temperature distribution of their disks than that of the
steady-state disk (Groot et al. 2001). This result, together with
our conclusions for the two SW Sex stars, 2MASS J01074282+4845188
(Khruzina et al. 2013) and J2256, imply that the non-steady-state
temperature distributions of the accretion disks is an additional
characteristic of the SW Sex-phenomenon.

(5) The ranges in the variability of the disk parameters of
J2256 in 2013--14 are: $a /a_0 \sim 0.27-0.33, \alpha_g \sim
0.48-0.58,T_{in} \sim 35000-42000$ K. For comparison, the
cataclysmic variables NZ Boo and V1239 Her show regular
oscillations of the disk parameters (radius, viscosity,
temperature, density) between the outbursts (steady states). Their
values of $\alpha_g$, $T_{in}$, and $\beta_d$ vary notably over
time even below 10 $P_{orb}$, whereas the disk radii change
slightly or are almost constant (Khruzina et al. 2015a, 2015b).

(6) The mean temperature $T_{wd}\approx$~23000 K of the J2256
white dwarf is in the range 19000--50000 K of the NLs (Warner,
1995).

(7) Most of the secondaries of the targets of NL, SW, UX types from
the last catalogue of CVs (Ritter $\&$ Kolb 2003, update RKcat7.22,
2014)  with periods in the range 0.16--0.33 d are K-type stars,
but there are also M-type dwarfs (XY Ari, V373 Pup, RW Tri, UX
UMa, LX Ser). The temperature of $T_{red} \sim$~3200 K of
J2256 means also an M-dwarf secondary, but this value is in the
range that corresponds to its binary period according to the spectral
type -- period relation for CVs of Knigge et al. (2011).

(8) The vast majority of CVs have mass ratios that are greater than
the limit $q = 1.2$ of a stable mass transfer of CVs (Schenker et
al. 1998). The derived mass ratio of J2256 of $q \sim 1.0\pm0.1$
is considerably lower. The review of the catalogue RKcat7.22
revealed two targets with small $q$: V363 Aur with \emph{q}=0.85 $\pm$ 0.05
and AC Cnc with \emph{q}=0.98 $\pm$ 0.04. But V363 Aur has an
orbital period of 0.3212 d and a late G star as the secondary
component, and AC Cnc has an orbital period of 0.3005 d and an
early K star; i.e., they have quite different configurations from that of
J2256. The system RW Tri seems more comparable to our target with
its period of P = 0.2286 d and an early M star secondary, but its
mass ratio $q = 1.3 \pm$ 0.4 cannot be considered as evidence of
an exception to the rule $q \geq$~1.2. Thus, the unusual
combination of parameters of J2256 turned out to have no analogue amongst
the known CVs.

(9) The fitting of the out-of-eclipse light curves of
J2256 required the emission of the hot line to be almost two times
bigger at phase 0.8 than at phase 0.2 (Tables 5--6). But this
means that the temperature of its windward side is lower than
that of the leeward side. This result was not expected from the
theory because the numerical 3D modelling of semi-detached
nonmagnetic binaries revealed that the stream from L1 causes
shockless interaction with the gaseous disk at stationary outflow
(Bisikalo et al. 1998, Matsuda et al. 1999, Makita et al. 2000).
As a result of the interaction of the stream with the coming parts
of the disk, an intensive shock wave arises at the lateral edges
of the stream. Its observational appearances on the leeward stream
side are equivalent to those of the hot spot on the disk.
Obviously, the shock wave direction determines the higher colour
temperature $T_{ww}$ of the matter at the windward stream part
than that of its leeward side $T_{lw}$. This is true for
the stationary state of the matter outflow.

The found peculiarity $T_{Iw} \geq T_{ww}$ of J2256
implies non-stationary overflow from the secondary component. We
have no plausible explanation of such a state during all of our
observational seasons, but possible reasons could be the variable
density of the gaseous stream and/or variable velocity of the
outflow (owing to the highly convective envelope of the cool
secondary). Some indications of this effect could be the observed
narrow, blue-shifted, transient absorption features in the Balmer
lines of DW UMa (Dhillon et al. 2013), which have been attributed to blobs of
material ejected from the system, possibly by a magnetic
propeller. Horne (1999) proposed the same explanation (magnetic
propeller) on the basis of the similarity in the spectral behaviour
of SW Sex stars and AE Aqr. Groot et al. (2001) went even further
by considering the possibility that the SW Sex stars are intermediate
polars at high inclination.

But we do not exclude the imperfections of the model as a possible
reason for the J2256 peculiarity $T_{Iw} \geq T_{ww}$.

\section{Analysis of the spectral data and SW Sex classification of J2256}

\begin{figure}
   \centering
   \includegraphics[width=9cm, angle=0]{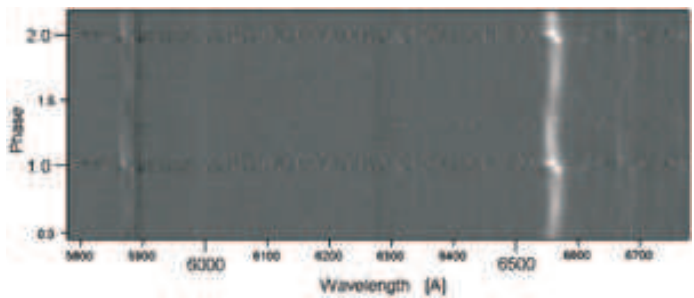}
   \caption{Medium-resolution trailed spectra of J2256 covering the orbital cycle}
   \label{Fig6}
   \end{figure}

   \begin{figure}
   \centering
   \includegraphics[width=9cm, angle=0]{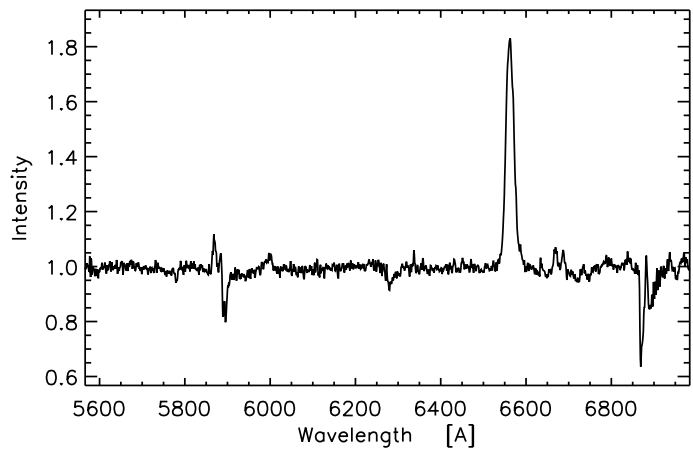}
   \caption{Averaged spectrum of J2256}
   \label{Fig6n}
   \end{figure}

  \begin{figure}
   \centering
   \includegraphics[width=8cm, angle=0]{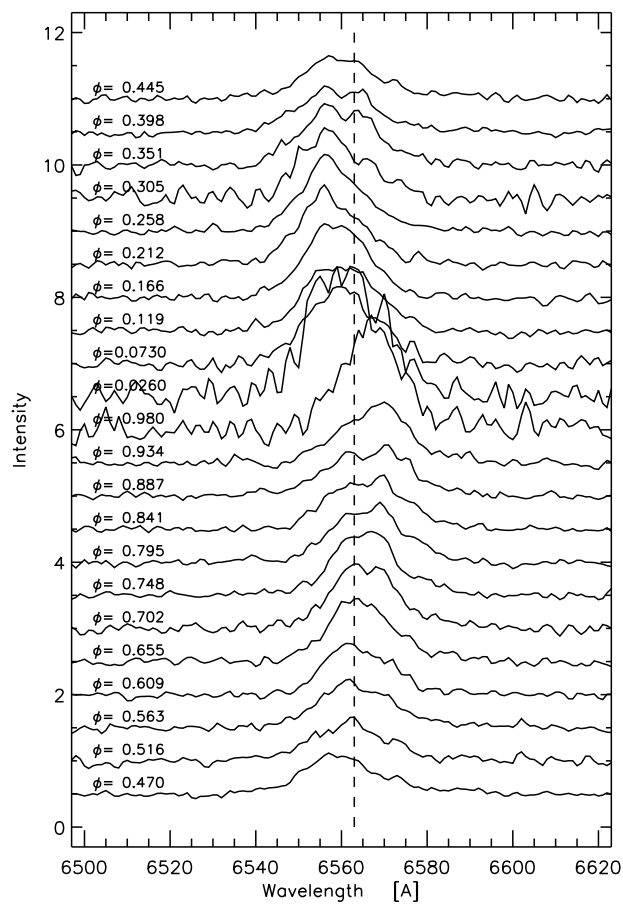}
   \caption{Variability of the H$\alpha$ line with the orbital phase (the laboratory wavelength is marked by a vertical dashed line)}
   \label{Fig7}
   \end{figure}

 \begin{figure}
   \centering
   \includegraphics[width=7cm, angle=0]{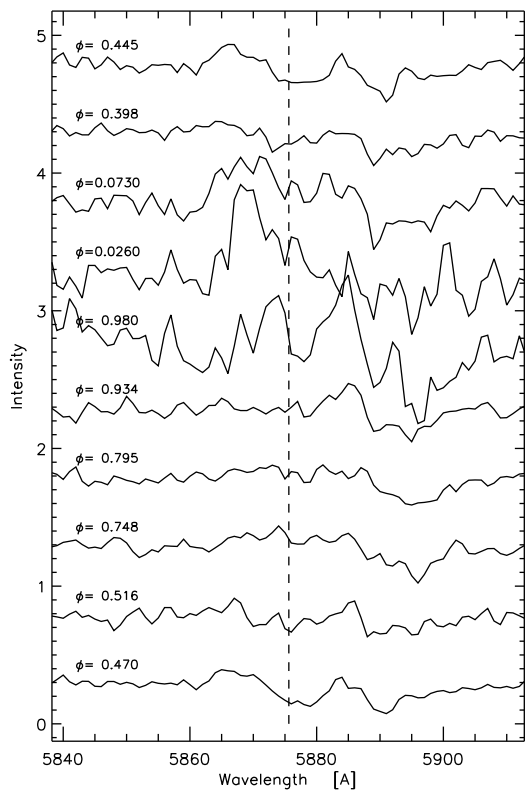}
   \caption{Illustration of the variability of the HeI~5876 line with the orbital phase}
   \label{Fig8}
   \end{figure}

Our medium-resolution spectra of J2256 allowed
semi-quantitative analysis and led to several conclusions.

(a) There are three emission lines in the observed range:
H$\alpha$, HeI~5876, and HeI~6678 (Fig.~6). The H$\alpha$ profiles
are single-peaked (similar to those of the SW Sex-subtype stars)
instead of double-peaked ones, as expected from near edge-on disks.
The profiles of the HeI~5876 and HeI~6678 emission lines look
double-peaked, at least at some phases (Fig. 6), but it is
difficult to determine their shape for this S/N. The
averaged spectrum of J2256 (Fig. 7) reveals more that
the HeI~6678 line is double-peaked. We found also weak two-peaked
emission lines of HeI in previous spectra of SW Sex stars: HeI
4471 and HeI 4921 of SW Sex (Dhillon et al. (1997); HeI 5785 and
HeI 6678 of SW Sex at phases 0.75--0.95 (Fig. 7 in Groot et al.
2011), weak HeI lines of UU Aqr (Hoard et al. 1998), and HeI 6678 of
AQ Men (Schmidtobreick et al. 2008).

Dhillon et al. (1997, 2013) supposed that the dominance of
the single-peaked line emission from the hot spot over the weak
double-peaked disk emission leads to the strange single-peaked
lines of the SW Sex-type spectra and their radial velocity curves
following the motion of the bright spot. The different amplitudes
of the radial velocities of the different emission lines are
another confirmation of this conclusion (Tovmassian et al. 2014).

The almost single-peaked emission profiles of J2256 imply
considerable hot-spot contribution. However, their rotational
disturbance around the mid-eclipse (Figs. 8-9) means that part of
the emission originates in the accretion disk. The secondary
star successively covers the disk regions that first approaches us
and then moves away from us.

(b) The averaged H$\alpha$ line of J2256 (Fig. 7) has an equivalent
width of 29 $\AA$. This value is less than that of SW Sex of 54
$\AA$ and close to that of DW UMa of 32 $\AA$ (Dhillon et al.
2013).

(c) Both the intensity and equivalent width (EW) of the H$\alpha$
line reveal phase variability (Fig. 10). The H$\alpha$ and
HeI~5876 emissions increase by a factor of 2 in the short phase
range 0.98--0.026, i.e. at the eclipse itself. In fact, the
emission lines become stronger with respect to the continuum
during the eclipse because the source of continuum radiation is
eclipsed (appears by the lower S/N of the spectra in Figs.
8--9). The abrupt increase in the H$\alpha$ emission at the
eclipse means that the H$\alpha$ emitting area is not eclipsed as
strongly as the continuum emitting area; i.e., it is located above
the orbital plane (Hoard et al. 2003). One can see similar
behaviour during the eclipse of the H$\alpha$ lines of HS
0129+2933, HS 0220+0603, and HS 0455+8513 (Rodriguez-Gil et al.
2007) as well as emission lines of UU Aqr (Hoard et al. 1998) and PG0818+513
(Thorstensen et al. 1991b).

It is tempting to suppose that the
increase in the emission in spectral lines at the eclipse is an
additional spectral property of the SW Sex phenomenon. But we do not want to forget that at the eclipse centre itself (phase 0.00), the
H$\alpha$ emission seems to decrease (see Fig. 7 of PG0027+260 in
Thorstensen et al. 1991a and Fig. 2 of IP Peg in Ishioka et al.
2004). Our spectra are not able to confirm the last effect because o
their low time resolution.

Although Figure 10 exhibits a decrease in the H$\alpha$
emission around phase 0.5, it is difficult to attribute it to the
effect ``phase 0.5 absorption'', which is very visible in some SW Sex
stars. Absorption features of the H$\beta$ and H$\gamma$ lines of
V1315 Aql, SW Sex, and DW UMa appear in the phase range 0.4-0.7
(Szkody $\&$ Piche 1990). Thorstensen et al. (1991a) found the
strong absorption at phase 0.5 for metal lines and HeI~6678 of the
SW Sex star PG0027+260. The trailed spectra of HS 0220+0603 and HS
0455+8315 clearly show the effect of the ``phase 0.5 absorption'' of
H$\alpha$, H$\beta$, and some HeI emission lines, but for HS
0129+2933 this effect is missing in the Balmer lines and presents in
the HeI emission lines (Fig. 10, Rodriguez-Gil et al. 2007). The
first deviation from the ``phase 0.5 absorption'' phenomenon was
UU Aqr with maximum absorption in the emission lines around
orbital phase 0.8 (Hoard et al. 1998).

Shallow absorption dips are visible on the top of the emission
H$\alpha$ lines of J2256 at phases 0.7--0.9 and 0.39-0.4 (Fig. 8),
which shift in phase with the wide emission profiles but with
considerably lower radial velocity. These dips are
rather individual peaks of the different emission sources than
some appearances of the ``phase 0.5 absorption'' effect.

(d) The trailed spectra (Fig. 6) reveal that the H$\alpha$ and
HeI~5876 emission lines of J2256 reveal S-wave wavelength shifts.
However, the smooth Doppler shifts of these lines are disturbed by
jumps (discontinuities) to the shorter wavelength around phase 0.0
(Figs. 6, 8, 9). The jump of the H$\alpha$ line amounts to more
than 10 $\AA$ while that of HeI~5876 seems bigger. The Doppler
jumps of the H$\alpha$ line around phase 0.0 were also observed
for other SW Sex stars (Fig. 9 in Rodriguez-Gil et al. 2007; Fig.
4 in Tovmassian et al. 2014), as well as for the lines H$\beta$ and
HeII 4686 of the eclipsing dwarf nova IP Peg (Ishioka et al.
2004).

Our medium-resolution spectra allow approximate estimations of the
radial velocities during the orbital cycle. For this aim we fitted
the H$\alpha$ lines of J2256 by Gausians (Thorstensen 2000).
Excluding the several outliers around phase 0.0, we
obtained semi-amplitude of the radial velocities of the emitting
material 210$\pm$50 km~s$^{-1}$ (Fig. 11), a typical value for SW
Sex type stars. The $\gamma$ velocity of J2256 is $-$85 km
s$^{-1}$.

(e) There is a phase discrepancy of $\sim$0.03 between the
photometric and spectral data of J2256 that is a typical property
of the SW Sex stars.

(f) The different widths of the emission lines of J2256 (Fig. 6)
mean that they originate in different disk regions with
different velocities and different temperatures.

(g) The absence of total disappearance of the emission
lines of J2256 means that they are formed above the disk and
probably originate in a vertically extended hot-spot halo.

The analysis of our medium-resolution spectra revealed that J2256
posses all spectral characteristics of the SW~Sex subtype of the
cataclysmic variables (defined by Thorstensen 1991, Rodriguez-Gil
et al. 2007, Schmidtobreick et al. 2009) apart from the 0.5
central absorption. The value $P_{\rm{orb}}=5.5$ h of J2256 is
slightly higher than the period range 3.0 -- 4.5 h of the known
SW~Sex stars, most of them tightly clustered just above the period
gap of CVs (Rodriguez-Gil et al. 2007). But there are members with
considerably longer periods (Bisol et al. 2012, Khruzina et al.
2013).

\begin{figure}
   \centering
   \includegraphics[width=7cm, angle=0]{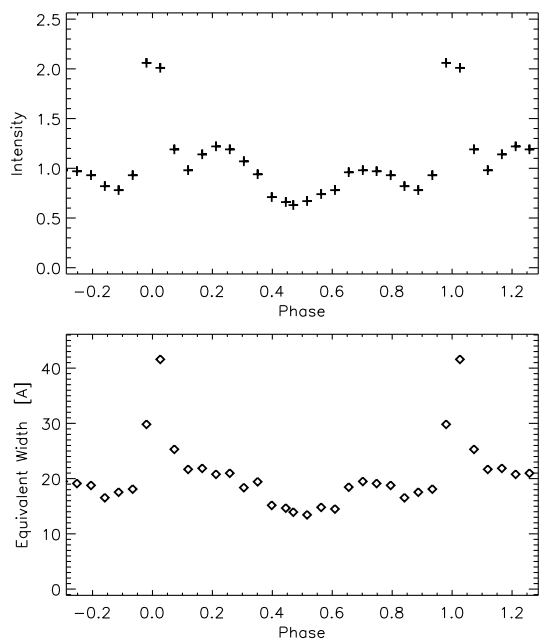}
   \caption{Phase variability of the intensity and EW of the H$\alpha$ line}
   \label{Fig14}
   \end{figure}

\begin{figure}
   \centering
   \includegraphics[width=7cm, angle=0]{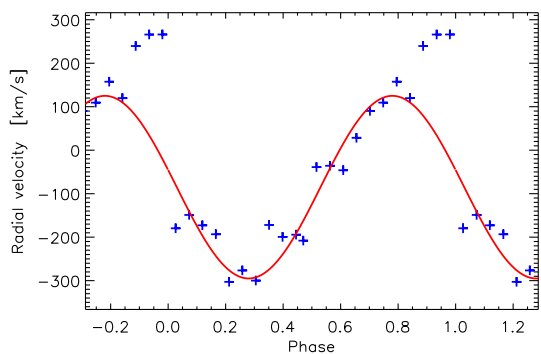}
   \caption{Variability in the radial velocities according to the measurement of the H$\alpha$ profiles}
   \label{Fig15}
   \end{figure}

\section{Conclusions}

The main results of the prolonged observations of the
newly discovered cataclysmic variable J2256, and their modelling
and analysis led to the following conclusions.
\begin{enumerate}

\item J2256 shows the deepest eclipse among the known
eclipsing nova-like variables.

\item The \emph{BVRI} light curves of the target were reproduced
well by a configuration consisting of a red dwarf and a white dwarf that is surrounded
by an accretion disk with a hot spot and a hot line.

\item The observed deep minimum of J2256 was reproduced by the
eclipse of a very bright, thin, small, but hot accretion
disk. The high temperature of its inner regions of
$\sim$38000--41000 K is typical of cataclysmic variables during
outburst.

\item The temperature distribution along the disk is flatter than
that of steady-state disk, which is typical of NLs. It was supposed that
the non-steady-state temperature distributions of the accretion
disks is an additional brand of the SW Sex-phenomenon.

\item The white dwarf of J2256 has a temperature around 22000 K,
while the secondary star is an M dwarf.

\item The derived mass ratio of J2256 of $q \sim 1.0\pm0.1$
is considerably lower than the limit $q = 1.2$ of the stable mass
transfer of CVs. The combination of the period, mass ratio, and
secondary temperature of J2256 turned out not to have an analogue amongst the known CVs.

\item The radiation of J2256 in the visual spectral range is
dominated by the emission of the accretion disk, whose relative
contribution is around 70 $\%$, followed by that of the secondary
star (around 20 $\%$) and hot line (around 10 $\%$).

\item We found an unusual temperature distribution of the hot line of
J2256: the temperature of its windward side is lower than that
of the leeward side.

\item The emission of J2256 in the lines H$\alpha$, HeI~5875, and
HeI~6678 increases considerably (twice) around the eclipse centre.
This effect is accompanied by large Doppler jumps of these
lines. The absence of eclipses of the emission lines and
their single-peaked profiles means that they originate mainly in a
vertically extended hot-spot halo.

\item The emission H$\alpha$ line reveals S-wave wavelength shifts
with semi-amplitude of around 210 km~s$^{-1}$ and phase lag of 0.03.

\item J2256 does not exhibit ``phase 0.5 absorption'' phenomena, which are
assumed to be a characteristic of the SW Sex stars.

\end{enumerate}

The high temperatures of the accretion disk of the newly
discovered cataclysmic star J2256 of SW~Sex subtype make it an
interesting target for follow-up observations in UV and X-ray
bands.

\begin{acknowledgements}
This study is supported by funds of the projects: RD 08-285/2015
of Shumen University, grant 14-02-00825 of the Russian Foundation
for fundamental Research and NSh-1675.2014.2 of the Programme of
State Support to Leading Scientific Schools. DK and DD gratefully
acknowledge observing grant support from the Institute of
Astronomy and Rozhen National Astronomical Observatory, Bulgarian
Academy of Sciences.

The authors are very grateful to the anonymous referee for
the valuable notes, advice, and suggestions.

This publication makes use of data products from the Two Micron
All Sky Survey, which is a joint project of the University of
Massachusetts and the Infrared Processing and Analysis
Center/California Institute of Technology, funded by the National
Aeronautics and Space Administration and the National Science
Foundation (Skrutskie et al. 2006). We have used also "The Guide
Star Catalogue, Version 2.3.2" (Lasker et al. 2008) and "The
USNO-B Catalog" (Monet et al. 2003). This research makes use of
the SIMBAD and Vizier data bases, operated at the CDS, Strasbourg,
France, and the NASA Astrophysics Data System Abstract Service.
The software {\sc Table curve 2D} is product of
Systat Software, Inc. (info-usa@systat.com, Corporate Headquarters).

\end{acknowledgements}

{}

\end{document}